\documentclass[conference]{IEEEtran}

\usepackage{xurl}
\usepackage[utf8]{inputenc} 
\usepackage[T1]{fontenc}
\usepackage{cite}
\usepackage{amsmath,amssymb}
\usepackage{graphicx}
\usepackage{booktabs}
\usepackage{array}
\usepackage{xcolor}         
\usepackage{url}
\usepackage[hidelinks]{hyperref}       

\title{From Prompts to Silicon: Scaling LLM-Assisted Chip Design with the AI Hardware Design League}

\author{
\IEEEauthorblockN{
Parsa Mirfasihi\IEEEauthorrefmark{1},
Harish Kumar Dharavath\IEEEauthorrefmark{1}
Muhtasim Alam Chowdhury\IEEEauthorrefmark{1},
Sujan Ghimire\IEEEauthorrefmark{1},\\
Banafsheh Saber Latibari\IEEEauthorrefmark{1},
Rozhin Yasaei\IEEEauthorrefmark{1},
Pratik Satam\IEEEauthorrefmark{2},
Soheil Salehi\IEEEauthorrefmark{1}
}
\IEEEauthorblockA{
\IEEEauthorrefmark{1}School of Electrical, Computing, and Software Engineering,
University of Arizona, Tucson, AZ 85721, USA\\
\IEEEauthorrefmark{2}College of Information Sciences,
University of Arizona, Tucson, AZ 85721, USA\\
Emails: \{parsamirfasihi, mmc7, sghimire, harrydhara16,
banafsheh,
yasaei, pratiksatam, ssalehi\}@arizona.edu
}
}

\begin{document}
\maketitle

\begin{abstract}
The semiconductor workforce shortage has renewed interest in making chip design accessible to students without conventional very-large-scale integration (VLSI) training. We present the AI Hardware Design League (AI-HDL), an international, project-based competition in which students use Large Language Models (LLMs) to generate and iteratively refine Verilog while simulation, synthesis, timing analysis, security review, and physical verification provide the engineering feedback needed to validate the result. Across two annual cycles, AI-HDL grew from 172 to 604 participants and from 40 to 108 teams, reaching 30 universities, six community colleges, and 19 countries. In the 2026 cohort, 68.7\% of participants reported no prior Verilog experience. Twenty-two teams completed the four design phases and reached final evaluation. Two division-winning designs completed physical sign-off in the SkyWater 130~nm (Sky130) process, passing design-rule checking and layout-versus-schematic verification with zero violations before being submitted for fabrication. These results show that an LLM-assisted, tool-grounded workflow can be deployed at substantial scale while also revealing that completion, rather than initial access, remains the principal challenge. AI-HDL materials and finalist artifacts are released to support replication and future study.
\end{abstract}

\vspace{1em}


\section{Introduction}
The Semiconductor Industry Association and Oxford Economics project that roughly 67,000 U.S. semiconductor positions could remain unfilled by 2030 at current degree-completion rates~\cite{sia2023}. At the same time, entry into hardware design remains demanding, as students must learn Hardware Description Languages (HDL), Register-Transfer-Level (RTL) reasoning, verification, and Electronic Design Automation (EDA) flows before they can produce a physical design. These requirements are especially difficult for learners at institutions without dedicated Very Large-Scale Integration (VLSI) chip design coursework or commercial EDA infrastructure.

Large Language Models (LLMs) and open-source EDA tools change this entry point. Recent work has shown that LLMs can generate useful Verilog from natural-language specifications~\cite{thakur2024verigen,liu2024rtllm,liu2023verilogeval}, while domain adaptation improves performance on chip-design tasks~\cite{liu2023chipnemo}. Chip-Chat further demonstrated that conversational hardware design can reach fabricated silicon~\cite{blocklove2023}. AutoChip further investigates iterative Verilog generation and repair using compiler and simulation feedback~\cite{Blocklove2025auto}. LLMs also support hardware weakness exploration, vulnerability-informed mitigation suggestions, and logic obfuscation for hardware IP protection~\cite{Ghimire2025_Hwrex, Ghimire2026canAgent}. Broader surveys situate these advances within an emerging landscape of LLM-assisted RTL generation, verification, debugging, and hardware security~\cite{ghimire2026largelanguagemodelsattentionbased}. In parallel, Yosys~\cite{yosys}, OpenLane~\cite{openlane}, OpenROAD~\cite{openroad}, and Sky130~\cite{sky130} have made synthesis and physical implementation available without proprietary tool licenses. These advances establish model and tool capability, but they do not by themselves show whether a large and heterogeneous student population can use them to traverse an end-to-end hardware-design workflow. 

AI Hardware Design League (AI-HDL) addresses this challenge as an educational training and workforce-development program. Teams use LLMs as the primary RTL-generating agent, retain prompt logs as part of their submissions, and progress through functional verification, Power-Performance-Area (PPA) optimization, security analysis, and physical implementation. The competition is intentionally phase-gated so that advancement depends on engineering artifacts rather than the apparent confidence of the model. Related initiatives emphasize specific objectives, such as RTL optimization or dataset development~\cite{ efabless2024challenge,iclad2026hackathon,iccd2025challenge,yu2024llm4hwdesign}, while AI-HDL addresses both and provides a sustained, mentored educational program spanning functional design, PPA optimization, security analysis and mitigation, and physical implementation, alongside a dataset of design artifacts and LLM interactions.

This paper makes three contributions. First, we describe a reproducible four-phase structure for LLM-assisted hardware design in which EDA feedback closes the loop between natural-language specification and verified implementation. Second, we report two annual deployments totaling 776 participants and quantify growth, prior experience, and progression through the program. Third, we report physical implementation results for two selected 2026 designs that reached clean SKY130 sign-off, demonstrating that the workflow can terminate in fabrication-ready hardware rather than only simulated RTL. The RTL source files, physical-design artifacts, and
sign-off reports for all eleven designs are publicly
available in the \href{https://github.com/prismlabarizona/AIHDL-2027/tree/master/AI_HDL_2026_Winner_Designs}
{AI-HDL repository}. 

\section{AI-HDL Program and Evaluation}
\label{sec:program}
\begin{figure*}[!t]
\centering
\includegraphics[width=0.9\linewidth]{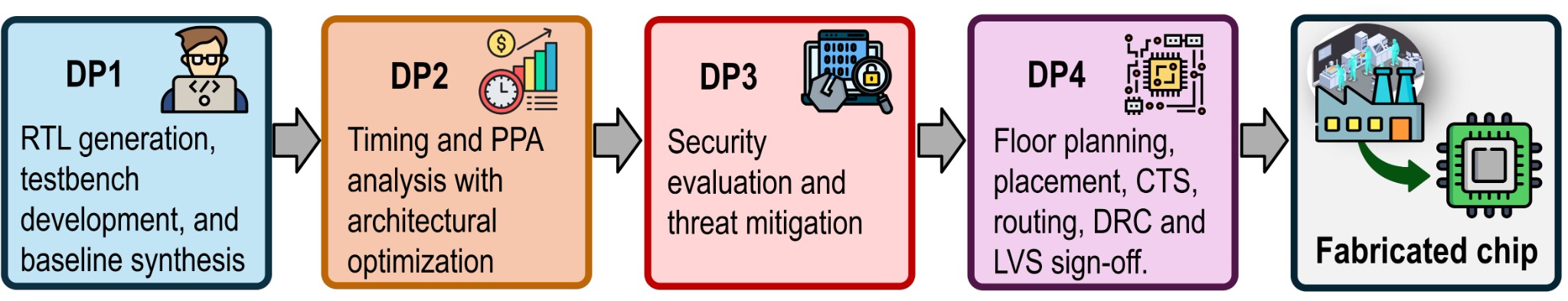}
\caption{AI-HDL Design Phases.}
\label{fig:Design_Phases}
\vspace{-2em}
\end{figure*}
\paragraph*{Competition structure}
The pilot cycle ran from October 2024 through March 2025 with 172 participants in 40 teams. Teams used AI tools to design a digital watch, prototype it on an FPGA, optimize it for PPA and security, and prepare it for tape-out. The second cycle ran from January through May 2026 with 604 participants in 108 teams and used a more demanding common RISC-V base design. Teams extended the processor with a custom peripheral and carried the design through four approximately month-long phases, as shown in Figure~\ref{fig:Design_Phases}: \textbf{DP1}, functional RTL, testbench development, and baseline synthesis. \textbf{DP2}, timing and PPA analysis followed by RTL or micro-architectural revision. \textbf{DP3}, security evaluation and threat mitigation. \textbf{DP4}, floor planning, placement, clock-tree synthesis, routing, and Design Rule Check (DRC) and Layout vs Schematic (LVS) sign-off. Each phase ended in a graded artifact, and teams that did not satisfy the phase rubric did not advance.

\paragraph*{LLM-assisted, tool-grounded workflow}
The central rule of AI-HDL is that teams generate the hardware by prompting an LLM and submit prompt logs with their design artifacts. Manual corrections are permitted but must be documented. Model output is never treated as evidence of correctness.
Functional behavior is checked through simulation using Icarus Verilog~\cite{icarus_verilog} and Verilator~\cite{verilator}, synthesis is performed with Yosys, and physical implementation is carried out with OpenLane on Sky130. Tool reports therefore provide an external corrective signal, which includes syntax errors, width mismatches, timing violations, and physical-design failures that must be resolved before a design can progress. Graduate mentors provide office hours and technical guidance, while advancement is determined by rubric-based evaluation of submitted artifacts.

\paragraph*{Industry partnership} Industry engagement is an important component of AI-HDL, providing resources and infrastructure that help scale the program, broaden access, and connect participants to the semiconductor ecosystem. 
Intel, Arm, and NVIDIA support participant recognition and awards, helping strengthen engagement and expand the program's visibility. AI-HDL is primarily built around open-source EDA software, while partnerships with Cadence and Synopsys provide participants with free access to industry-standard EDA tools, removing a major practical barrier to hands-on chip design. ChipMango serves as a curriculum-development partner and offers free beginner- and intermediate-level VLSI design certificate programs to AI-HDL participants. Finally, ChipFoundry serves as the program's tapeout partner, providing a pathway for selected designs to progress from physical signoff to fabricated silicon.
Together, these partnerships create an end-to-end ecosystem that connects education and competition with professional design tools, industry-recognized skills, and real-world semiconductor fabrication.

\paragraph*{Evaluation data}
We summarize the program using registration records, onboarding surveys, phase-gate outcomes, finalist evaluations, and physical implementation reports. Prior experience is self-reported. The pilot survey was optional and therefore has item-specific denominators; the 2026 Verilog-experience result reported below uses the full cohort denominator of 604. Because AI-HDL is an operating program rather than a controlled intervention, the results characterize reach, progression, and engineering outcomes, and they do not isolate the causal contribution of the LLM from mentoring, competition incentives, templates, or the open-source toolchain.

\section{Scale, Access, and Progression}
\label{sec:results}

\begin{figure*}[t]
\centering
\includegraphics[width=0.9\linewidth]{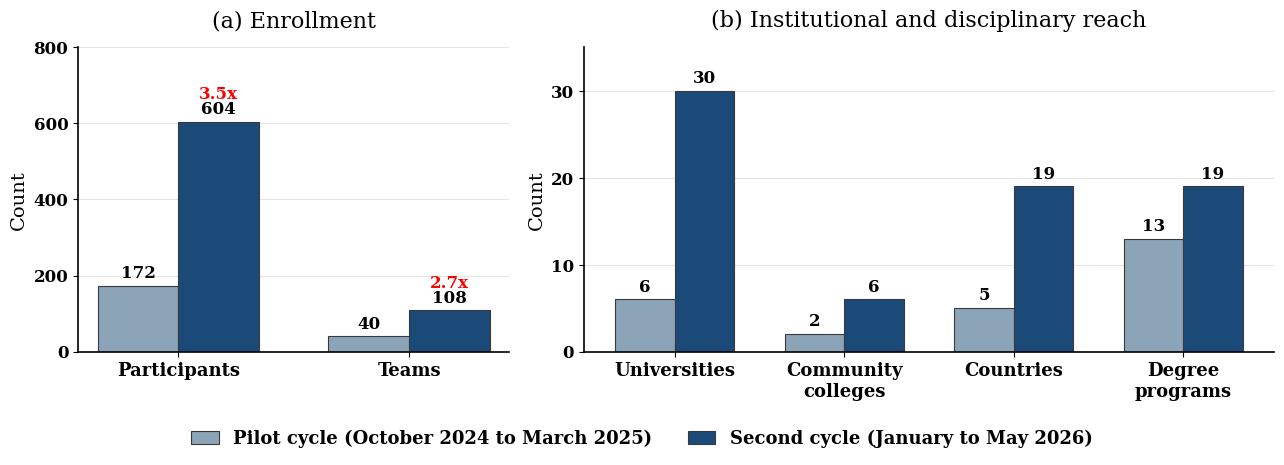}
\caption{Growth from the 2025 pilot to the 2026 cycle. Participation increased from 172 to 604 and teams from 40 to 108, while the program expanded to 30 universities, six community colleges, 19 countries, and 19 reported degree programs. 
}
\label{fig:growth}
\end{figure*}

\paragraph*{Scale and reach}
Across the two cycles, AI-HDL enrolled 776 participants. As shown in Figure~\ref{fig:growth}, enrollment increased by 3.5$\times$ and the number of teams by 2.7$\times$ between the pilot and second cycle. Institutional reach expanded from six to 30 universities and from 2 to 6 community colleges, while geographic representation grew from 5 to 19 countries. Participants also reported 19 degree programs in 2026, extending well beyond electrical and computer engineering. The expansion across institutions and disciplines is important for the program's workforce objective because it indicates that growth was not limited to a larger cohort at a single host institution.

\paragraph*{Prior experience}
The 2026 cohort was dominated by learners with limited HDL background, where 415 of 604 participants (68.7\%) reported no prior Verilog experience, while another 150 reported only beginner-level experience. Thus, 565 of 604 participants (93.5\%) entered with either no experience or beginner-level experience in Verilog. Prior familiarity with generative AI and LLMs was also limited, with 166 participants (27.5\%) reporting no prior experience with generative AI or LLMs and 278 (46.0\%) reporting only beginner-level experience, so 444 of 604 (73.5\%) entered with little or no working familiarity with LLMs, while 133 (22.0\%) described themselves as knowledgeable and 27 (4.5\%) as expert. 
These results indicate that AI-HDL reached participants
for whom RTL development was not yet an established skill.

\begin{figure}[!t]
\centering
\includegraphics[width=0.50\textwidth]{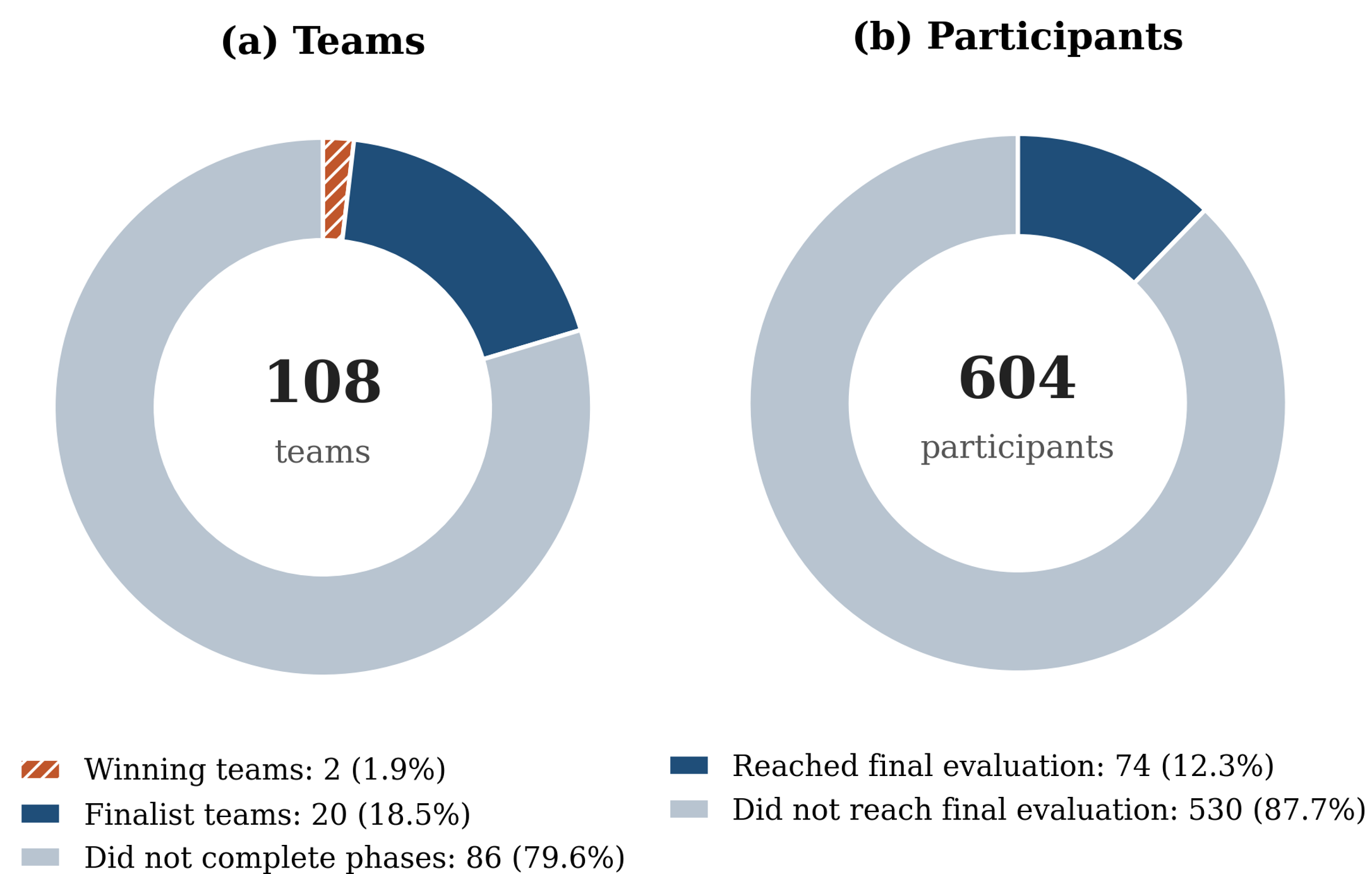}
\caption{Progression in the 2026 cycle. Of 108 registered teams, 22 completed all four design phases and reached final evaluation, representing 20.4\% of registered teams and 74 participants. Two selected designs proceeded to physical sign-off and fabrication submission.}
\label{fig:funnel}
\end{figure}

\paragraph*{Progression}
Figure~\ref{fig:funnel} reports the completion of the final design phase rather than only the winning teams. Of 108 registered teams, 22 completed the four phases and reached final evaluation, corresponding to 20.4\% of registered teams, and the teams represented 74 participants. Open registration makes initial participation intentionally low-friction, whereas later phases require functional verification, timing closure, security review, and physical implementation. The resulting gap between entry and completion is therefore a program outcome rather than a statistic to hide. AI-HDL scaled access substantially, but sustaining teams through an end-to-end hardware flow remains the primary challenge for future cycles.

\section{Physical Implementation Outcomes}
\label{sec:silicon}

The 2026 challenge used a common small RISC-V base while allowing teams to develop custom peripherals. Two division-winning designs were selected for final physical implementation, which were an SPI controller with CRC-16 packet checking and an AES-128 accelerator integrated with the RISC-V system. Both were implemented in Sky130 and passed DRC and LVS with zero violations before being sent for fabrication. Table~\ref{tab:silicon} reports a compact subset of the post-layout results. The values should be interpreted as two implementation case studies rather than a head-to-head comparison because the blocks serve different functions and were optimized under different design objectives.

\begin{table}[htbp]
\centering
\caption{Selected post-layout results for the two 2026 division-winning designs at 1.8~V. Power values are post-layout estimates from the implementation flow.}
\label{tab:silicon}
\small
\begin{tabular}{@{}lcc@{}}
\toprule
\textbf{Metric} & \textbf{SPI + CRC-16} & \textbf{AES-128} \\
\midrule
Core area ($\mu$m$^2$) & 37,488 & 149,082 \\
Standard cells & 2,392 & 6,492 \\
Clock frequency (MHz) & 71.43 & 57.10 \\
Estimated total power & 2.29~$\mu$W & 5.68~mW \\
DRC/LVS violations & 0 & 0 \\
\bottomrule
\end{tabular}
\end{table}

\subsection{Shared-Chip Layout and Selected Design Results}
\label{sec:shared_chip_layout}
\begin{figure}[!t]
    \centering  \includegraphics[width=\columnwidth,height=\columnwidth,keepaspectratio]{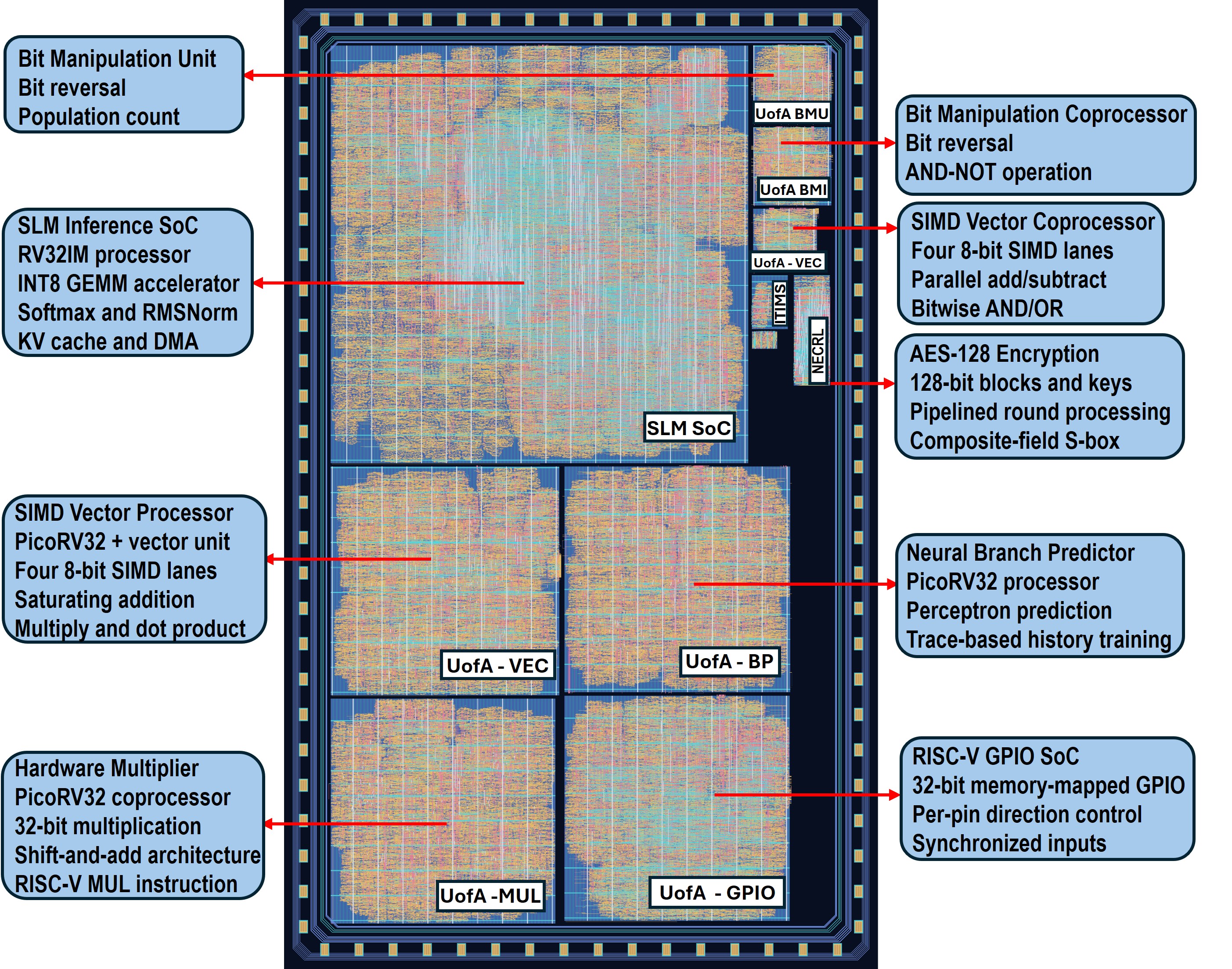}
    \caption{Composite layout visualization of eleven designs submitted for fabrication through the AI-HDL tapeout effort. All designs completed physical sign-off, including two AI-HDL 2026 division-winning student designs and nine additional designs finalized by mentors and professional engineers.}
    \label{fig:ai_hdl_chip_layout}
\end{figure}

Figure~\ref{fig:ai_hdl_chip_layout} presents a composite layout visualization of separate designs prepared for the AI-HDL tapeout with ChipFoundry, including the SLM SoC and RISC-V-based designs with custom peripherals. Among these, the two best overall designs are an SPI controller with CRC-16 packet checking and an AES-128 encryption accelerator, each integrated with its respective RISC-V system. As summarized in Table~\ref{tab:silicon}, the SPI design occupies 37,488~$\mu$m$^2$ of core area and operates at 71.43~MHz, while the AES-128 design occupies 149,082~$\mu$m$^2$ and operates at 57.10~MHz. Both achieved zero DRC and LVS violations in Sky130. These results describe two distinct implementations with different functions and design objectives.

These implementations provide a concrete endpoint for the competition workflow. The result is not that an LLM autonomously designed a correct chip. Rather, students used an LLM to generate and revise RTL while simulators, synthesis, timing analysis, and physical verification determined whether each revision was acceptable. All eleven designs successfully completed physical sign-off for fabrication, including the two AI-HDL division-winning designs and nine additional designs finalized by mentors and professional engineers. This demonstrates the ability of the AI-HDL ecosystem to support physically realizable hardware designs.

\section{AI-HDL 2026 Timeline}
\begin{figure}[t]
    \centering  \includegraphics[width=\columnwidth,height=\columnwidth,keepaspectratio]{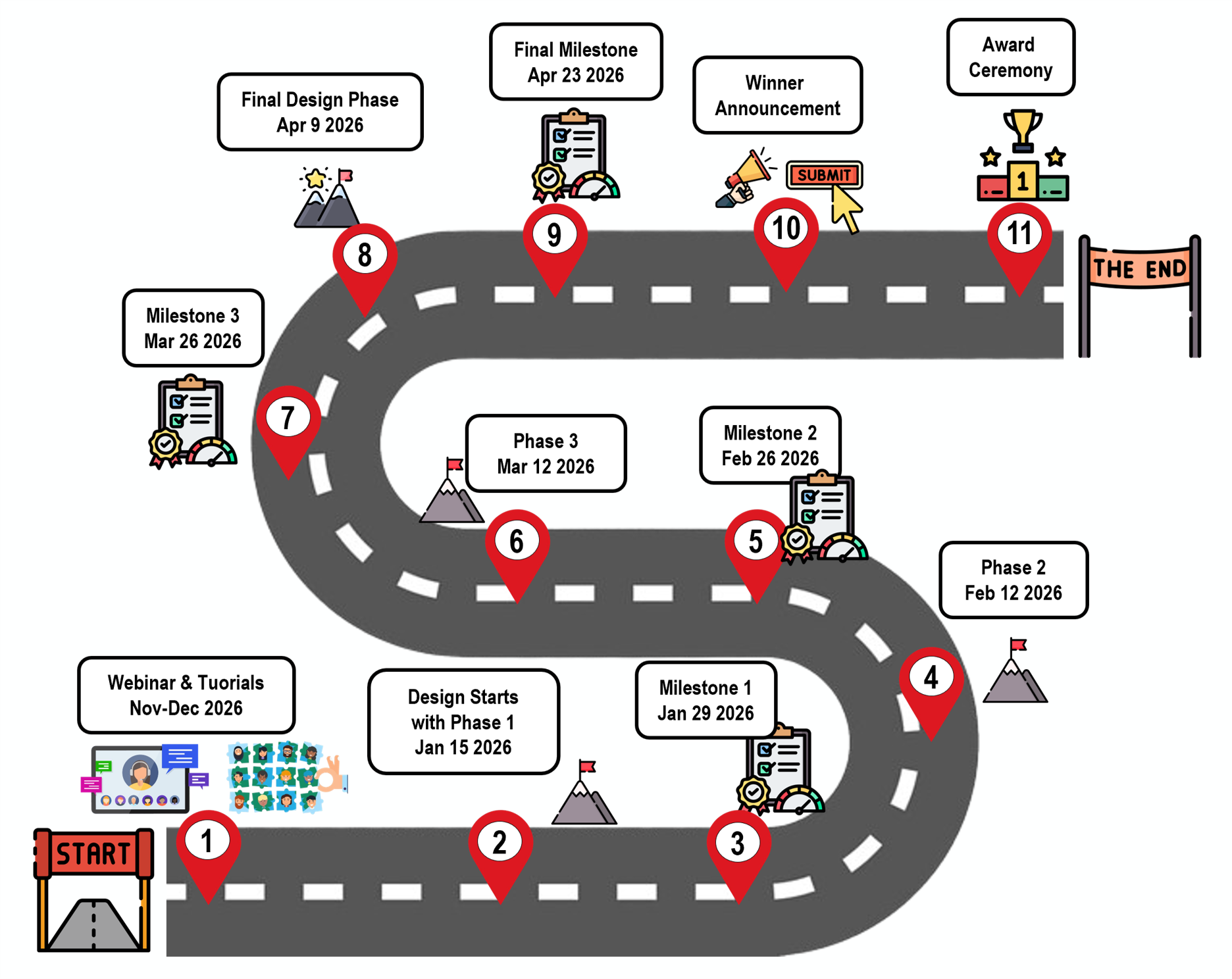}
    \caption{The Competition Timeline of AI-HDL 2026.}
    \label{fig:schedule}
    \vspace{-1.5em}
\end{figure}

The AI-HDL 2026 competition follows a structured multi-stage schedule spanning from the pre-competition preparation period through the final award ceremony. Before the official design phase begins, participants attend a series of webinars and tutorials from November to December to become familiar with the competition objectives, design flow, tools, and submission requirements. Registration also closes during this period. The main competition begins in January, when contestants start Phase 1 and progressively develop and refine their hardware designs through the DP1–DP4 stages in January, February, March, and April, respectively. Each phase is accompanied by intermediate milestones that allow teams to demonstrate progress and prepare for the subsequent stage of the competition. The final design phase concludes in April, followed by the final milestone submission and evaluation of the completed designs. After the submissions are reviewed and ranked according to the competition criteria, the winning teams are announced in May, followed by the award ceremony. The complete sequence of competition phases, milestones, and major events is illustrated in Figure~\ref{fig:schedule}.

\section{AI-HDL 2026 Award Ceremony}
The award ceremony is held at the University of Arizona. The ceremony will give an overview of the AI-HDL 2026 Contest, introducing the Industry panel for a talk (AI for Chip Design) followed by QnA and interaction with the students, and finally announcing the contest results, and present the awards to the winners. 

\section{Discussion and Limitations}
\label{sec:discussion}
Together, the two cycles demonstrate that an LLM-assisted, tool-grounded hardware-design program can operate at an international scale, engage participants with limited prior HDL experience, and produce verified engineering artifacts. The results also show a clear boundary. Registration growth is much larger than completion growth, and only 20.4\% of 2026 teams reached final evaluation. Lowering the barrier to a first RTL implementation therefore does not remove the difficulty of timing closure, verification, security analysis, or physical design.

Several limitations constrain stronger claims. AI-HDL has no non-LLM control group, so the effect of prompting cannot be separated from mentoring, phase gates, templates, or fabrication incentives. Prior experience was self-reported, and the competition
did not administer pre- and post-program assessments
to directly measure learning gains. Future cycles could adopt formative assessments informed by prior engineering-laboratory studies ~\cite{Baiyun2017Elevating} and personalized AI tutoring inspired by PRISM~\cite{Lin2026}. Teams were free to choose their LLM, and prompt logs document submitted interactions but cannot prove that no unreported manual authoring occurred. Finally, the physical results come from two selected designs on a single open process node and should not be treated as representative of every participating team or of larger industrial designs. Future evaluation should track phase-level attrition more closely, add direct learning measures, and use the accumulated prompt logs to study how novice design strategies change as EDA feedback becomes more demanding. 

\section{Conclusion}
AI-HDL provides a scalable model for introducing students to end-to-end chip design through a combination of generative AI, structured project phases, mentoring, and open-source EDA. Across two cycles, 776 participants entered the program. The 2026 cohort reached 30 universities and 19 countries, and 68.7\% reported no prior Verilog experience. Twenty-two teams completed all four phases, and two selected designs reached clean Sky130 physical sign-off and were sent for fabrication. The central lesson is therefore two-sided: LLMs can provide a practical entry point into hardware design when their outputs are grounded in engineering tools, but access alone does not guarantee completion. The remaining challenge is to scale the verification, physical-design support, and instructional structure needed to carry a broader fraction of novice teams from an initial prompt to verified silicon.


\bibliographystyle{IEEEtran}
\bibliography{References_ISCAS_2027}

\end{document}